\documentclass{rQUF2e}

\usepackage{epstopdf}% To incorporate .eps illustrations using PDFLaTeX, etc.
\usepackage{subfigure}% Support for small, `sub' figures and tables
\usepackage{subeqnarray}
\usepackage{amsmath,amssymb,amsthm,mathrsfs,amsfonts,dsfont} 
\usepackage{color}

\theoremstyle{plain}

\newcommand{\erf}{\textrm{erf}}
\newcommand{\sgn}{\textrm{sign}}

\theoremstyle{definition}

\theoremstyle{remark}

\begin{document}

%\jvol{00} \jnum{00} \jyear{2014} \jmonth{October}

\title{Market impact with multi-timescale liquidity}

\author{M. BENZAQUEN$^{\ast}$$\dag$${\ddag}$\thanks{$^\ast$Corresponding author.
Email: michael.benzaquen@polytechnique.edu} and J.-P. BOUCHAUD${\ddag}$\\
\affil{$\dag$ Ladhyx, UMR CNRS 7646, \'Ecole Polytechnique,
91128 Palaiseau Cedex, France\\
$\ddag$  Capital Fund Management, 23 rue de l'Universit\'e, 75007, Paris, France} \received{\today} 
}

\maketitle

\begin{abstract}
We present an extended version of the recently proposed ``LLOB'' model for the dynamics of latent liquidity in financial markets. By allowing for finite cancellation and deposition rates within a continuous reaction-diffusion setup, we account for finite memory effects on the dynamics of the latent order book. We compute in particular the finite memory corrections to the square root impact law, as well as the impact decay and the permanent impact of a meta-order. { The latter is found to be linear  in the traded volume and independent of the trading rate, as dictated by no-arbitrage arguments}. In addition, we consider the case of a spectrum of cancellation and deposition rates, which allows us to obtain a square root impact law for moderate participation rates, as observed empirically. Our multi-scale framework also provides an alternative solution to the so-called price diffusivity puzzle in the presence of a long-range correlated order flow.
\end{abstract}


%\begin{classcode}Please provide at least one JEL Classification code\end{classcode}

\section*{Introduction}
\label{intro}

Understanding the price formation mechanisms is undoubtably among the most exciting challenges of modern finance. \emph{Market impact} refers to the way market participants' actions mechanically affect prices. Significant progress has been made in this direction during the past decades \cite{Hasbrouk2007,bouchaud2008markets,weber2005order,Bouchaud_impact_2010}. A notable breakthrough was the empirical discovery that the aggregate price impact of a meta-order\footnote{A ``meta-order'' (or parent order) is a bundle of orders
corresponding to a single trading decision. A meta-order is typically traded incrementally through a sequence of child orders.} is a concave function (approximately square-root) of its size $Q$ \cite{Grinold,Almgren2005,Toth2011,Donier2015}. In the recent past, so called ``latent'' order book models \cite{Toth2011,mastromatteo2014agent,MPRL,DonierLLOB} have proven to be a fruitful framework to theoretically address the question of market impact, among others.\\

As a precise mathematical incarnation of the latent order book idea, the zero-intelligence LLOB model of Donier \textit{et al.} \cite{DonierLLOB} was successful at providing a theoretical underpinning to the square root impact law. The LLOB model is based on a continuous mean field setting, that leads to a set of reaction-diffusion equations for the dynamics of the latent bid and ask volume densities. In the infinite memory limit (where the agents intentions, unless executed, stay in the latent book forever and there are no arrivals of new intentions), the latent order book becomes exactly linear and impact exactly square-root. Furthermore, this assumption leads to zero permanent impact of uninformed trades, and an inverse square root decay of impact as a function of time. 
While the LLOB model is fully consistent mathematically, it suffers from at least two major difficulties when confronted with micro-data. First, a strict square-root law is only recovered in the limit where the execution rate $m_0$ of the meta-order is larger than the normal execution rate $J$ of the market itself -- whereas most meta-order impact data is in the 
opposite limit $m_0 \lesssim 0.1 J$. Second, the theoretical inverse square-root impact decay is too fast and leads to significant short time mean-reversion effects, not observed in real prices. \\

The aim of the present paper is to show that introducing different timescales for the renewal of liquidity allows one to cure both the above deficiencies. In view of the way financial markets operate, this step is very natural: agents are indeed expected to display a broad spectrum of timescales, from low frequency institutional investors to High Frequency Traders (HFT). We show that provided the execution rate $m_0$ is large compared to the low-frequency flow, but small compared to $J$, the impact of a meta-order crosses over from a linear behaviour at very small $Q$ to a square-root law in a regime of $Q$s that can be made compatible with empirical data. We show that in the presence of a continuous, power-law distribution of memory times, the temporal decay of impact can be tuned to reconcile persistent order flow with diffusive price dynamics (often referred to as the \emph{diffusivity puzzle})  \cite{bouchaud2008markets,bouchaud2004fluctuations,Lillo2004}.  We argue that the permanent impact of uninformed trades is fixed by the slowest liquidity memory time, beyond which mean-reversion effects disappear. Interestingly, the permanent impact is found to be linear { in the executed volume $Q$ and independent of the trading rate}, as dictated by no-arbitrage arguments.\\

Our paper is organized as follows. We first recall the LLOB model of \cite{DonierLLOB} in Section~\ref{llobrecall}. We then explore in Section~\ref{fincandep} the implications of finite cancellation and deposition rates (finite memory) in the reaction-diffusion equations, notably regarding permanent impact (Section~\ref{permimp}). We generalize the reaction-diffusion model to account for several deposition and cancellation rates. In particular, we analyse in Section~\ref{multifsec} the simplified case of a market with two sorts of agents: long memory agents with vanishing deposition and cancellation rates, and short memory high frequency agents (somehow playing the role of market makers). Finally, we consider in Section~\ref{densnusec} the more realistic case of a continuous distribution of cancellation and deposition rates and show that such a framework provides an alternative way to solve the diffusivity puzzle (see \cite{BenzaquenFLOB}) by adjusting the distribution of cancellation and deposition rates. Many details of the calculations are provided in the Appendices.

\section{Locally linear order book model}
\label{llobrecall}

We here briefly recall the main ingredients of the locally linear order book (LLOB) model as presented by Donier \textit{et al.} \cite{DonierLLOB}. In the continuous ``hydrodynamic'' limit we define the latent volume densities of limit orders in the order book: $\varphi_{\mathrm{b}}(x,t)$ (bid side) and $\varphi_{\mathrm{a}}(x,t)$ (ask side) at price $x$ and time $t$. The latter obey the following set of partial differential equations:
\begin{subeqnarray}
\partial_t \varphi_{\mathrm{b}} &=& D\partial_{xx}\varphi_{\mathrm{b}} -\nu\varphi_{\mathrm{b}} + \lambda \Theta(x_t-x) - R_\mathrm{ab}(x) \slabel{goveqsnl1}\\
\partial_t \varphi_{\mathrm{a}} &=& D\partial_{xx}\varphi_{\mathrm{a}} -\nu\varphi_{\mathrm{a}}  + \lambda \Theta(x-x_t) - R_\mathrm{ab}(x)\  ,\quad \ 
\slabel{goveqsnl2}
\end{subeqnarray}
where the different contributions on the right hand side respectively signify (from left to right): heterogeneous reassessments of agents intentions with diffusivity $D$ (diffusion terms), cancellations with rate $\nu$ (death terms), arrivals of new intentions with intensity $\lambda$ (deposition terms), and matching of buy/sell intentions (reaction terms). The price $x_t$ is conventionally defined through the equation $ \varphi_{\mathrm{b}}(x_t,t)= \varphi_{\mathrm{a}}(x_t,t)$. 
\begin{figure}[t!]
\begin{center}
\resizebox{0.48\columnwidth}{!}{  \includegraphics{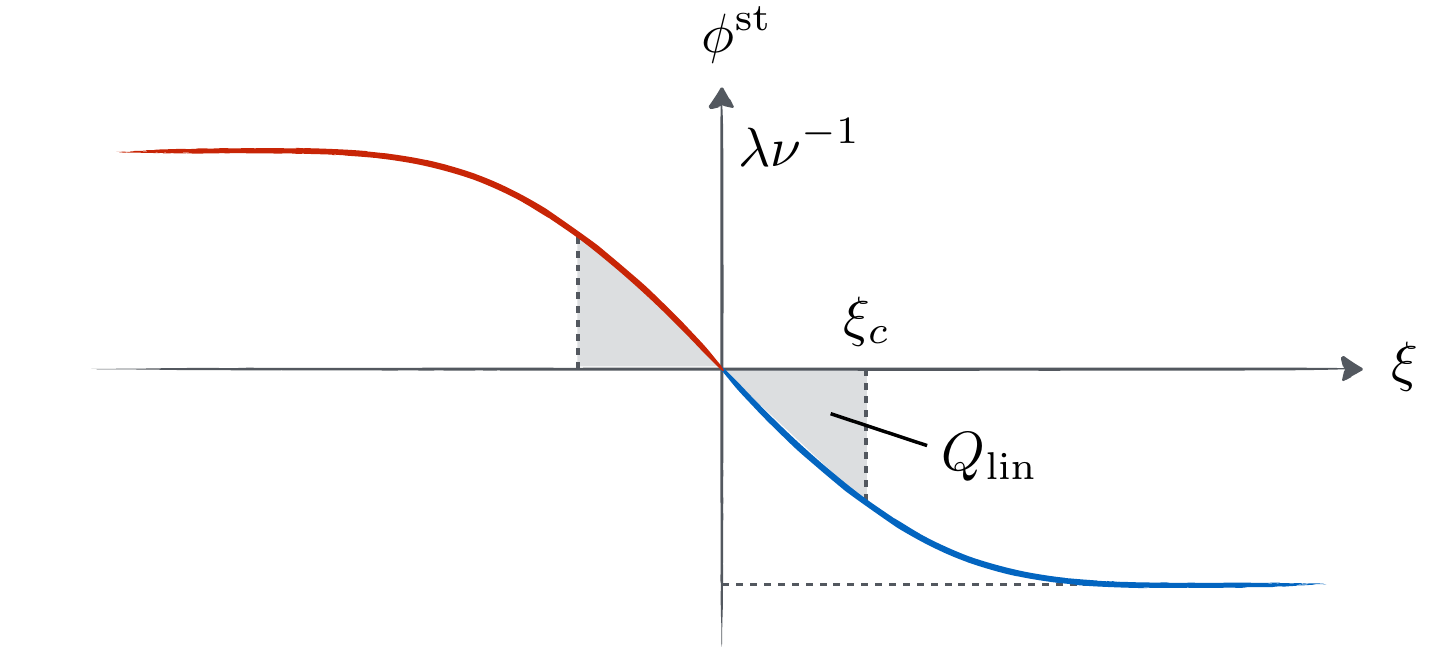}}
\end{center}
\caption{Stationary order book $\phi^\mathrm{st}(\xi)$ as computed by Donier \emph{et al.} \cite{DonierLLOB}. The linear approximation holds up to $\xi_{\mathrm c}=\sqrt{D\nu^{-1}}$ and the volume $Q_\mathrm{lin.}$ of the grey triangles is of order $Q_\mathrm{lin.}:=\mathcal{L}\xi_\mathrm{c}^2= J \nu^{-1}$.}
\label{Obstat}      
\end{figure}
The non-linearity arising from the reaction term in Eqs. \eqref{goveqsnl1} and \eqref{goveqsnl2} can be abstracted away by defining  $ \phi(x,t) = \varphi_{\textrm b}(x, t) - \varphi_{\textrm a}(x, t)$, which solves:
\begin{eqnarray}
\partial_t \phi &=& D \partial_{xx} \phi -\nu\phi + s(x,t) \ ,\label{firsteqsrc}
\end{eqnarray}
where the source term reads $s(x,t) = \lambda \,\sgn (x_t-x)$ and the price $x_t$ is defined as the solution of 
\begin{eqnarray}
\phi(x_t,t) &=& 0 \ .  \label{priceeq}
\end{eqnarray}
Setting $\xi=x-x_t$, the stationary order book can easily be obtained as: $\phi^\mathrm{st}(\xi)=-({\lambda}/{\nu}) \, \sgn(\xi)  [1-\exp(-|\xi|/\xi_{\mathrm c})]$ where $\xi_{\mathrm c}=\sqrt{D\nu^{-1}}$ denotes the typical length scale below which the order book can be considered to be linear: $\phi^\mathrm{st}(\xi) \approx -\mathcal L \xi$ (see Fig.~\ref{Obstat}). The slope $\mathcal L := \lambda/\sqrt{\nu D}$ defines the {\it liquidity} of the market, from which the total execution rate $J$ can be computed since:
\begin{eqnarray}
J := \left. \partial_\xi \phi^\mathrm{st}(\xi)  \right|_{\xi=0} = D \mathcal{L}.
\end{eqnarray}
Donier \emph{et al.} \cite{DonierLLOB} focussed on the \emph{infinite memory} limit, namely $\nu, \lambda \rightarrow 0$ while keeping $\mathcal L \sim \lambda {\nu}^{-1/2}$ constant, such that the latent order book becomes exactly linear since in that limit $\xi_{\mathrm c} \to \infty$. This limit considerably simplifies the mathematical analysis, in particular concerning the impact of a meta-order. An important remark must however be introduced at this point: although the limit $\nu \to 0$ is taken in \cite{DonierLLOB}, it is assumed that 
the latent order book is still able to reach its stationary state $\phi^\mathrm{st}(\xi)$ before a meta-order is introduced. In other words, the limit $\nu \to 0$ is understood in a way such that the starting time of the meta-order is large compared to $\nu^{-1}$.    

\section{Price trajectories with finite cancellation and deposition rates}
\label{fincandep}

As mentioned in the introduction we here wish to explore the effects of non-vanishing cancellation and deposition rates, or said differently the behaviour of market impact for executiong times larger than $\nu^{-1}$. The general solution of Eq.~\eqref{firsteqsrc} is given by:
\begin{eqnarray}
\phi(x,t) &=& \left( \mathcal G_\nu \star \phi_0\right)(x,t) + \int \text d y\int_0^\infty \text d \tau\, \mathcal G_\nu(x-y,t-\tau) s(y,\tau) \ , \label{convol}
\end{eqnarray}
where $\phi_0(x) =\phi(x,0)$ denotes the initial condition, and where $\mathcal G_\nu (x,t) = e^{-\nu t}\mathcal G (x,t)$ with $\mathcal G$ the diffusion kernel:
\begin{eqnarray}
\mathcal G(x,t) &=& \Theta(t)  \frac{e^{-\frac{x^2}{4Dt}}}{\sqrt{4\pi Dt}} \ .
\end{eqnarray}
Following Donier \emph{et al.} \cite{DonierLLOB}, we introduce a buy (sell) meta-order as an extra point-like source of buy (sell) particles with intensity rate $m_t$ such that the source term in Eq.~\eqref{firsteqsrc} becomes: $s(x,t) = m_t \delta(x-x_t)\cdot \mathds{1}_{[0,T]} +\lambda \,\sgn (x_t-x)$, where $T$ denotes the time horizon of the execution. In all the following we shall focus on buy meta-orders -- without loss of generality since within the present framework everything is perfectly symmetric. Performing the integral over space in Eq.~\eqref{convol} and setting $\phi_0(x)=\phi^{\mathrm{st}}(x)$ yields:
\begin{eqnarray}
\phi(x,t) &=&  \phi^\mathrm{st}(x)e^{-\nu t} + \int_0^{\min (t,T)} \text d \tau\, m_\tau \mathcal G_\nu(x-x_\tau,t-\tau)    -\lambda\int_0^{t } \text d \tau \,  \erf\left[  \frac{x-x_\tau}{\sqrt{4D(t-\tau)}}  \right] e^{-\nu(t-\tau)} \ .\label{mastereq}
\end{eqnarray}
The equation for price, \eqref{priceeq}, is not analytically  tractable in the general case, but different interesting limit cases can be investigated. In particular, focussing on the case of constant participation rates $m_t = m_0$, one may consider:
\begin{itemize}
\item (\emph{i}) Small participation rate $m_0\ll J$ \emph{vs} large participation rate $m_0\gg J$.\medskip

\item (\emph{ii}) Fast execution $\nu T\ll 1$ (the particules in the book are barely renewed during the meta-order execution) \emph{vs} slow execution $\nu T\gg 1$ (the particles in the book are completely renewed, and the memory of the initial state has been lost).\medskip

\item (\emph{iii}) Small meta-order volumes $Q:=m_0 T\ll Q_\mathrm{lin.}$ (for which the linear approximation of the stationary book is appropriate, see Fig.~\ref{Obstat}) \emph{vs} large volumes $Q \gg Q_\mathrm{lin.}$ (for which the linear approximation is no longer valid).
\end{itemize}

So in principle, one has to consider $2^3 = 8$ possible limit regimes. However, some regimes are mutually exclusives so that only 6 of them remain. A convenient way to summarize the results obtained for each of the limit cases mentioned above is to expand the price trajectory $x_t$ up to first order in $\sqrt{\nu}$ as:\footnote{Note that working at constant $\mathcal L$ implies  $\lambda=O\big(\sqrt{\nu}\big)$.}
\begin{eqnarray}
x_t &=& \alpha \left[ z_t^0+\sqrt{\nu} z_t^1+O(\nu)\right] \ ,\label{alphaz0z1}
\end{eqnarray}
where $z_t^0$ and $z_t^1$ denote respectively the 0th order and 1st order contributions. Table 1 gathers the results for fast execution ($\nu T\ll 1$) and small meta-order volumes 
($Q \ll Q_\mathrm{lin.}$). Note that the leading correction term $z_t^1$ is negative, i.e. the extra incoming flux of limit orders acts to lower the impact of the meta-order, see Fig.~\ref{pricetraj}. 
The price trajectory for slow execution and/or large meta-order volumes, on the other hand, simply reads: 
\begin{eqnarray}
x_t &=& \frac{m_0 \nu}{\lambda} t \ . 
\end{eqnarray}
The corresponding calculations and explanations are given in Appendix A.
\medskip

\begin{table*}[t!]
    \centering
\resizebox{1\textwidth}{!}{  \includegraphics{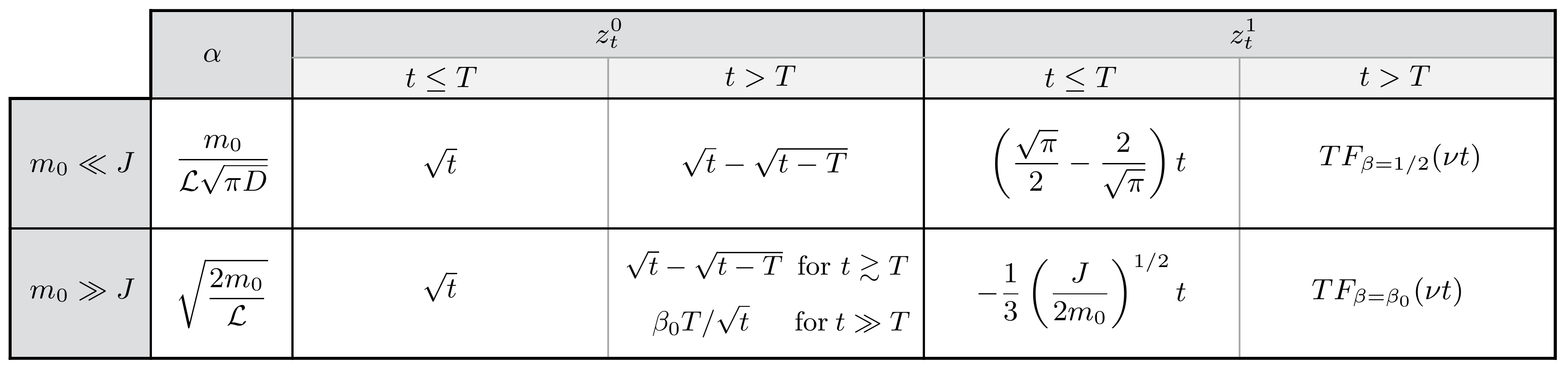}}\medskip
\caption{Price trajectories for different impact regimes (see Eq.~\eqref{alphaz0z1}). We set $\beta_0 := { \frac12 }\left[{{m_0}/(2\pi J})\right]^{1/2}$.}
\label{tableimpact}     
\end{table*}

\section{Permanent impact as a finite memory effect}
\label{permimp}

As mentioned in the introduction, the impact relaxation following the execution is an equally important question. We here compute the impact decay after a meta-order execution. In the limit of small cancellation rates, we look for a scaling solution of the form $z^1_t= T F(\nu t)$ (see Eq.~\eqref{alphaz0z1}) where $F$ is a dimensionless function. We consider the case where $\nu T \ll 1$ and $Q \ll Q_\mathrm{lin.}$. Long after the end of the execution of the meta-order, i.e. when $t\gg T$,  Eq.~\eqref{priceeq} together with Eqs.~\eqref{mastereq} and \eqref{alphaz0z1} becomes (to leading order): 
\begin{eqnarray}
0&=& -\frac{\lambda\alpha T}{\sqrt{D}}F(\nu t)e^{-\nu t} - 2\lambda\alpha\int_0^{t } \text d \tau \,    \frac{z_t^0-z_\tau^0}{\sqrt{4\pi D(t-\tau)}}  e^{-\nu(t-\tau)} \nonumber \\
&& \quad \quad \quad \quad \quad \quad \quad \quad \quad \quad \quad \quad \quad \quad \quad\quad \quad \quad   -2\lambda \alpha T\sqrt{\nu} \int_0^t  \text d \tau \,    \frac{F(\nu t)-F(\nu \tau )}{\sqrt{4\pi D(t-\tau)}}  e^{-\nu(t-\tau)}
\ .\label{}
\end{eqnarray}
Letting $u = \nu t$ and $z_t^0 = \beta/\sqrt{u}$ (see Table \ref{tableimpact}) yields:
\begin{eqnarray}
0&=&\sqrt{\pi}e^{-u} F(u) + \beta \int_0^{u } \text d v \,    \frac{\sqrt{v} -\sqrt{u}}{\sqrt{uv(u-v)}}  e^{v-u} +  \int_0^{u } \text d v \,    \frac{F(u)-F(v) }{\sqrt{u-v}}  e^{v-u}
\ .\label{intequ}
\end{eqnarray}
Finally seeking $F$ asymptotically of the form $F(u) = F_\infty +Bu^{-\gamma}+Cu^{-\delta}e^{-u}$ 
one can show that:
\begin{eqnarray}
F(u)&=&F_\infty -\frac{\beta}{\sqrt{u}}\left[1-e^{-u}\right] \qquad (u \gg 1)\   ,\label{Fu}
\end{eqnarray}
with the permanent component given by $F_\infty = \beta\sqrt{\pi}$, where $\beta$ depends on the fast/slow nature of the execution (see Table \ref{tableimpact}).\\

\begin{figure}[t!]
\begin{center}
\resizebox{0.5\columnwidth}{!}{\includegraphics{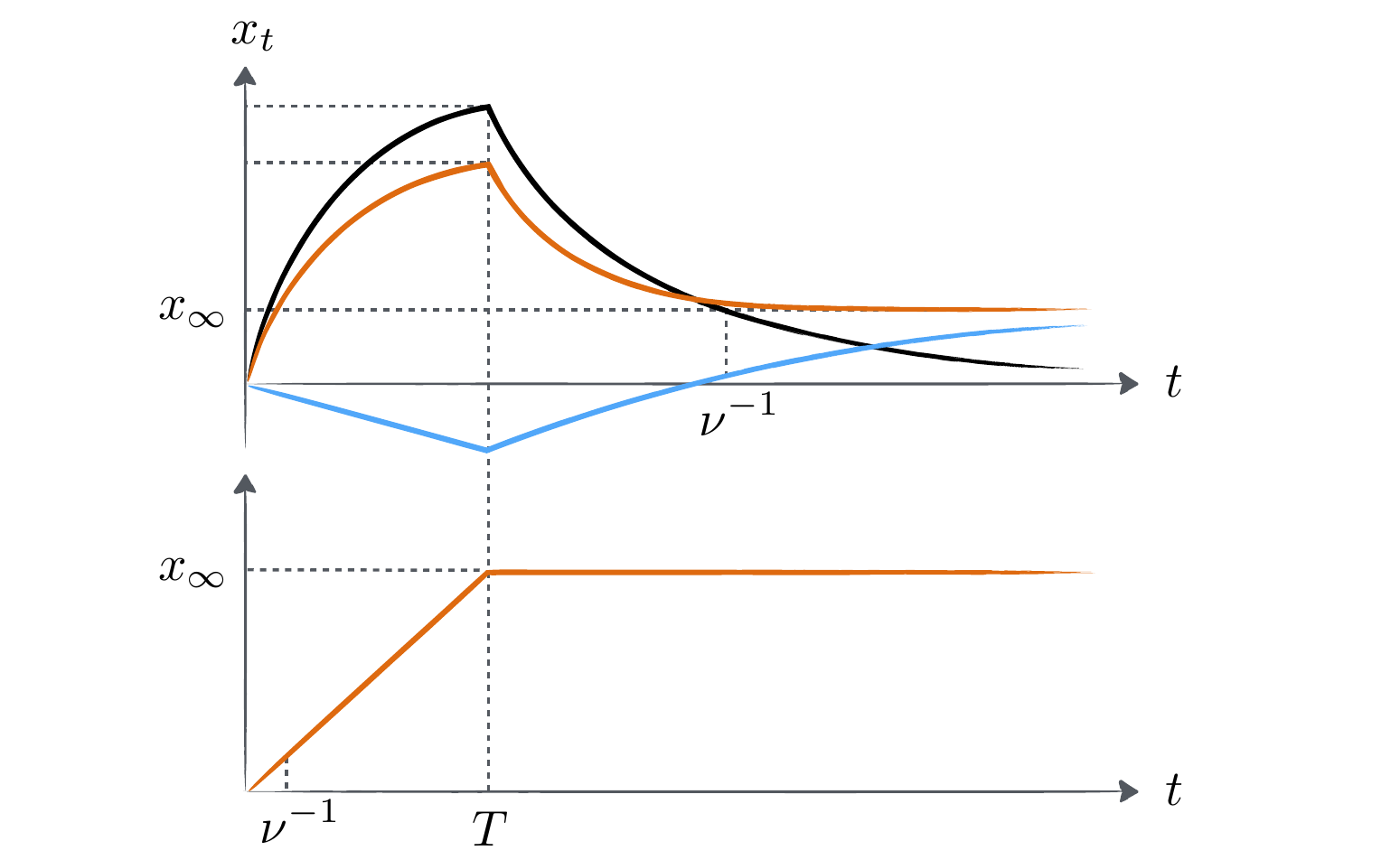}}
\end{center}
\caption{Top graph: Price trajectory during and after a buy meta-order execution for $\nu T\ll 1$. (Black curve) 0th order result from \cite{DonierLLOB}. (Orange curve) 1st order result. (Blue curve) 1st order correction (see Eq.~\eqref{alphaz0z1}). Bottom graph: Price trajectory for $\nu T\gg1 $. Note that the $x$-axis is not to scale since $\nu^{-1}\ll$ (resp. $\gg$) $ T$.}
\label{pricetraj}      
\end{figure}

Injecting the solution for $F(u)$ in Eq.~\eqref{alphaz0z1}, and taking the limit of large times, one finds that the $t^{-1/2}$ decay of the 0th order term is exactly compensated 
by the $\beta u^{-1/2}$ term coming from $F(u)$, showing that the asymptotic value of the impact, given by $I_\infty= \alpha \sqrt{\nu} { T} F_\infty$, is reached exponentially fast as $\nu t \to \infty$ (see Fig.~\ref{pricetraj}). This result can be interpreted as follows. At the end of execution (when the peak impact is reached), the impact starts decaying towards zero in a slow power law fashion (see \cite{DonierLLOB}) until approximately $t \sim \nu^{-1}$, beyond which all memory is lost (since the book has been globally renewed). Impact cannot decay anymore, since the previous reference price has been forgotten. 
Note that in the limit of large meta-order volumes and/or slow executions, all memory is already lost at the end of the execution and the permanent impact trivially matches the peak impact (see Fig.~\ref{pricetraj}).\\

An important remark is in order here. Using Table \ref{tableimpact}, one finds that $I_\infty {  = \frac12} \xi_c (Q/Q_\mathrm{lin.})$ in both the small and large participation regime. In other words, we find that the permanent impact is {\it linear} in the executed volume $Q$, as dictated by no-arbitrage arguments { \cite{huberman2004price,gatheral2010no}} and compatible with the classical Kyle framework { \cite{kyle1985continuous}}.

\section{Impact with fast and slow traders}
\label{multifsec}

\subsection{Set up of the problem}

As stated in the introduction, one major issue in the impact results of the LLOB model as presented by Donier \textit{et al.} \cite{DonierLLOB} is the following. Empirically, the impact
of meta-orders is only weakly dependent on the participation rate $m_0/J$ (see e.g. \cite{Toth2011}). The corresponding \emph{square root law} is commonly written as:
\begin{eqnarray}
I_Q &:=& \langle x_T \rangle = Y \sigma \sqrt{\frac Q V} \ , \label{empiricalimp}
\end{eqnarray}
where $\sigma$ is the daily volatility, $V$ is the daily traded volume, and $Y$ is a numerical constant of order unity. Note that $I_Q$ only depends on the total volume of the meta-order $Q=m_0 T$, and not on $m_0$ (or equivalently on the time $T$). \\

\begin{figure}[t!]
\begin{center}
\resizebox{0.48\columnwidth}{!}{  \includegraphics{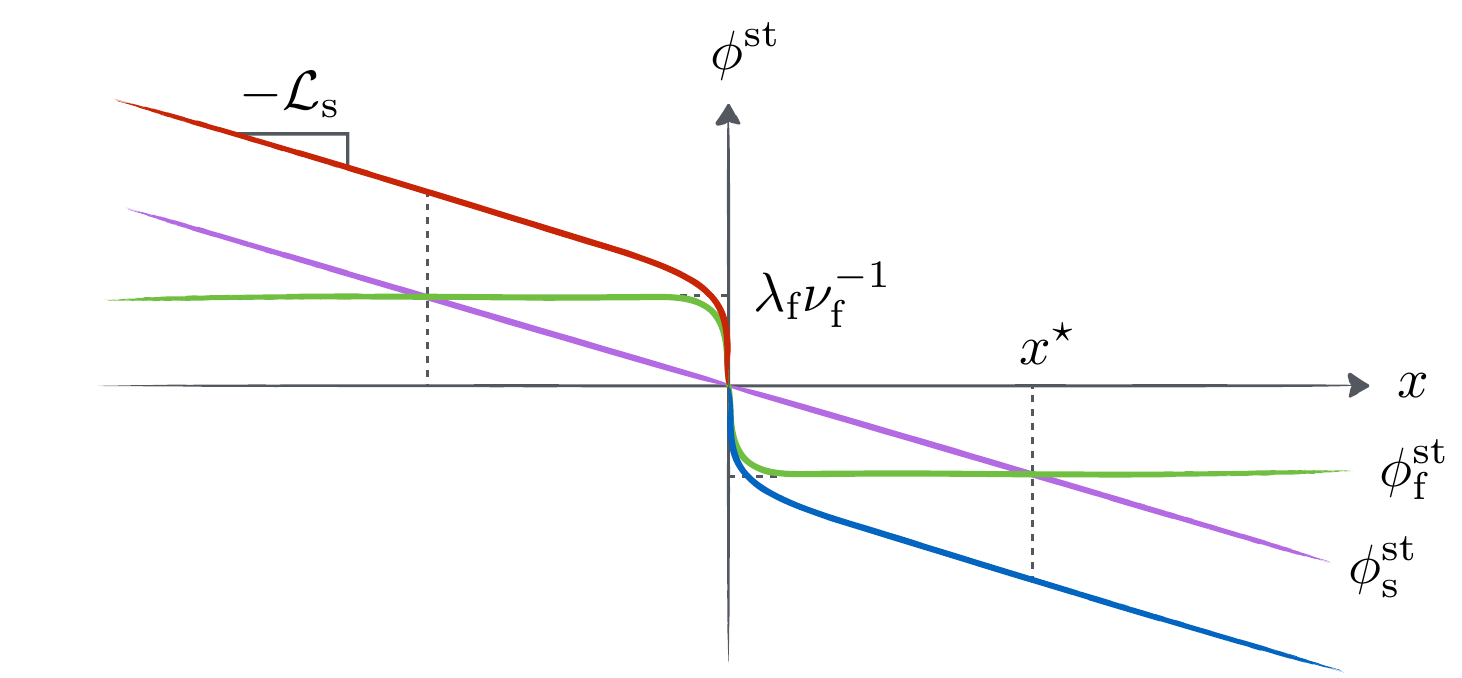}}
\end{center}
\caption{Stationary double-frequency order book $\phi^\textrm{st}(x)=\phi_{\text{s}}^\textrm{st}(x)$ (purple) $+\ \phi_{\text{f}}^\textrm{st}(x)$ (green) (see Section \ref{multifsec}).}
\label{Obstat_multi}      
\end{figure}

As one can check from Table \ref{tableimpact}, the independence of impact on $m_0$ only holds in the large participation rate limit ($m_0\gg J$). However, most investors choose to operate in the opposite limit of small participation rates $m_0 \ll J$, and all the available data is indeed restricted to $m_0/J \lesssim 0.1$. Here we offer a possible way out of this conundrum. The intuition is that the total market turnover $J$ is dominated by high frequency traders/market makers, whereas resistance to slow meta-orders can only be provided by slow participants on the other side of the book. 
More precisely, consider that only two sorts of agents co-exist in the market (see Section~\ref{densnusec} for a continuous range of frequencies): 
\begin{enumerate}
\item Slow agents with vanishing cancellation and deposition rates: $\nu_{\text{s}} T \rightarrow 0$,  while keeping the corresponding liquidity $\mathcal L_{\text{s}}:= \lambda_{\text{s}}/\sqrt{\nu_{\text{s}} D}$ finite; and 
\item Fast agents with large cancellation and deposition rates, $\nu_{\text{f}} T \gg 1$, such that $\mathcal L_{\text{f}}:= \lambda_{\text{f}}/\sqrt{\nu_{\text{f}} D} \gg \mathcal L_{\text{s}}$.
\end{enumerate}
The system of partial differential equations to solve now reads:
\begin{subeqnarray}
\partial_t \phi_{\text{s}} &=& D \partial_{xx} \phi_{\text{s}} -\nu_{\text{s}}\phi_{\text{s}} +s_{\text{s}}(x,t) \slabel{phi1}
\\
\partial_t \phi_{\text{f}} &=& D \partial_{xx} \phi_{\text{f}} -\nu_{\text{f}}\phi_{\text{f}} +s_{\text{f}}(x,t) \ ,\slabel{phi2}
\end{subeqnarray}
where $s_k(x,t) = \lambda_k \,\sgn (x_{kt}-x) +  m_{kt}\delta(x-x_{kt}) $, together with the conditions:
\begin{eqnarray}
m_{\text{s}t}+ m_{\text{f}t} &=& m_0 \label{ratesequal} \\
x_{\text{s}t}=x_{\text{f}t} &=& x_t  \label{priceequal}\ .
\end{eqnarray}
Equation~\eqref{ratesequal} means that the meta-order is executed against slow and fast agents, respectively contributing to the rates $m_{\text{s}t}$ and $m_{\text{f}t}$. Equation~\eqref{priceequal} simply means that there is a unique transaction price, the same for slow and for fast agents.
The total order book volume density is then  given by $\phi =\phi_{\text{s}}+\phi_{\text{f}}$. In particular, in the limit of slow/fast agents discussed above the stationary order book is given by the sum of $\phi_{\text{s}}^\mathrm{st}(x) \approx -\mathcal L_{\text{s}} x$  and $\phi_{\text{f}}^\mathrm{st}(x) \approx - (\lambda_{\text{f}}/\nu_{\text{f}})\sgn(x)$ (see Fig.~\ref{Obstat_multi}). The total transaction rate now reads 
\begin{eqnarray}
J  =  \ D \left|\partial_x\left[ \phi_{\text{s}}^\mathrm{st}+\phi_{\text{f}}^\mathrm{st}\right]\right|_{x=0}=J_{\text{s}}+J_{\text{f}},
\end{eqnarray}
where  $J_{\text{f}} \gg J_{\text{s}}$ (which notably implies that $J \approx J_{\text{f}}$). 

%%%%%%%%%
%%%%%%%%%

\subsection{From linear to square-root impact}

\begin{figure}[t!]
\begin{center}
\resizebox{0.39\columnwidth}{!}{  \includegraphics{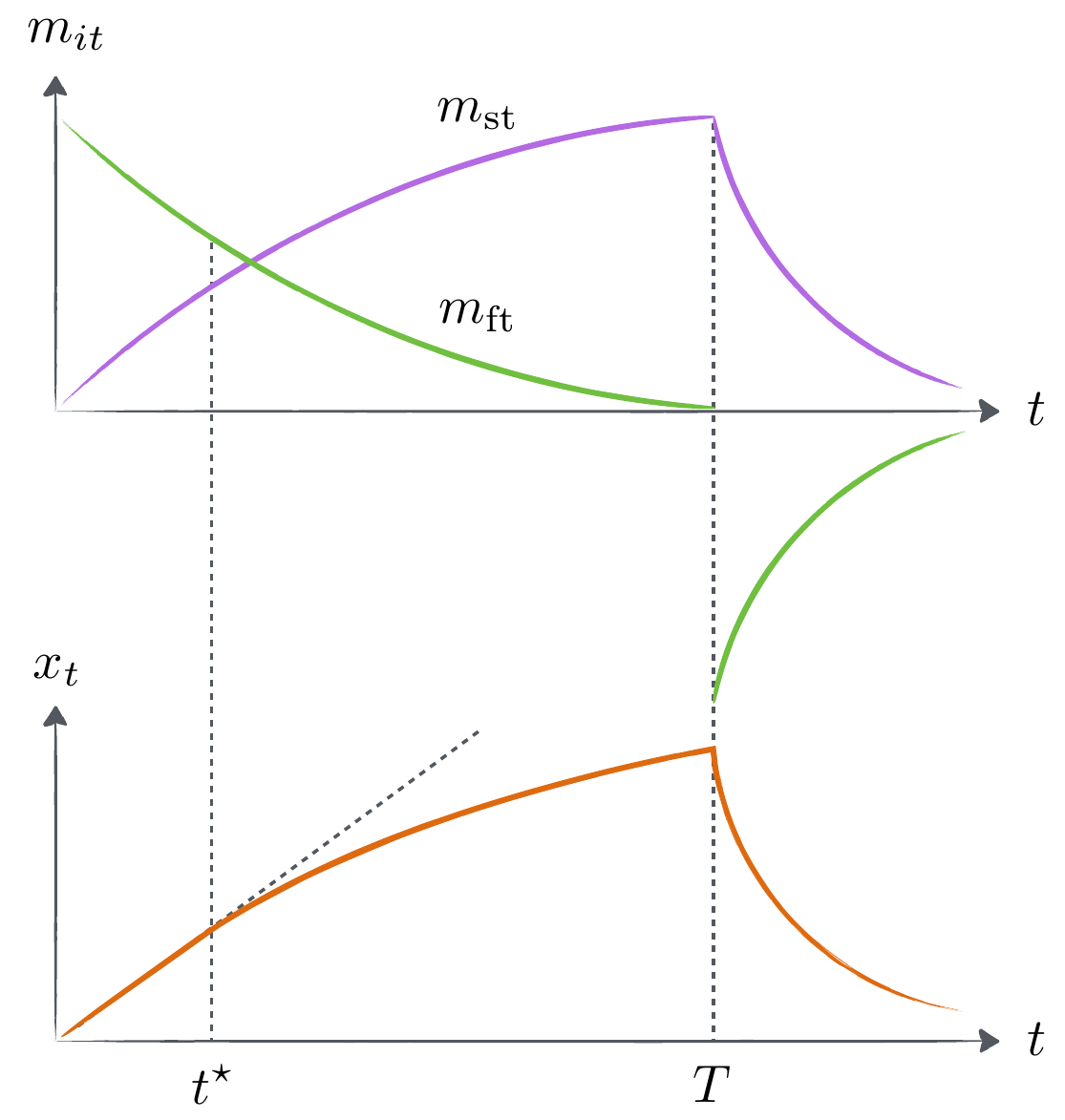}}
\end{center}
\caption{Execution rates $m_{it}$ (top) and price trajectory (bottom) within the double-frequency order book model (see Section \ref{multifsec}).}
\label{pricetraj_multi}      
\end{figure}

We now focus on the regime where the meta-order intensity is large compared to the average transaction rate of slow traders, but small compared to the total transaction rate of the market, to wit: $J_{\text{s}} \ll m_0 \ll J$. In this limit Eqs.~\eqref{phi1} and \eqref{phi2}, together with the corresponding price setting equations $\phi_k(x_{kt},t) \equiv 0$ yield (see Appendix B):
\begin{subeqnarray}
x_{\text{s}t} &=&\left(\frac2{\mathcal L_{\text{s}}} \int_0^t \textrm d \tau \,  m_{\text{s}\tau}  \right)^{1/2} \slabel{x1tspec}\\
x_{\text{f}t} &=& \frac{\nu_{\text{f}}}{\lambda_{\text{f}}} \int_0^t \textrm d \tau \,  m_{\text{f}\tau}  \ .  \slabel{x2tspec} \label{x1tx2t}
\end{subeqnarray}
Differentiating Eq.~\eqref{priceequal} with respect to time together with Eqs.~\eqref{x1tx2t} and using Eq.~\eqref{ratesequal} yields:
\begin{eqnarray}
m_{\text{f}t} &=&\frac{m_0}{\sqrt{1+\frac{t}{t^\star}}}, \quad \text{with} \quad t^\star:=\frac{1}{2\nu_{\text{f}}} \frac{J_{\text{f}}^2}{J_{\text{s}}m_0}, \label{m2}
\end{eqnarray}
and $m_{\text{s}t}=m_0-m_{\text{f}t}$.  Equation \eqref{m2} indicates that most of the incoming meta-order is executed against the rapid agents for $t < t^\star$ but the slow agents then take over for $t>t^\star$ (see Fig.~\ref{pricetraj_multi}).
The resulting price trajectory reads:
\begin{eqnarray}
x_{t} &=&\frac{\lambda_{\text{f}}}{\mathcal L_{\text{s}} \nu_{\text{f}}}\left(\sqrt{1+\frac{t}{t^\star}}-1\right) \, , \label{ptrajmultif}
\end{eqnarray}
which crosses over from a linear regime when $t \ll t^\star$ to a square root regime for $t \gg t^\star$ (see Fig.~\ref{pricetraj_multi}). For a meta-order of volume $Q$ executed during a time interval $T$, the corresponding impact is linear in $Q$ when $T < t^\star$ and square-root (with $I_Q$ independent of $m_0$) when $T > t^\star$. This last regime takes place when $Q > m_0 t^\star$, which can be rewritten as:
\begin{eqnarray}
\frac{Q}{V_{\text{d}}} > \frac{1}{\nu_{\text{f}} T_{\text{d}}} \frac{J}{J_{\text{s}}},
\end{eqnarray}
where $V_{\text{d}}$ is the total daily volume and $T_{\text{d}}$ is one trading day. Numerically, with a HFT cancellation rate of -- say -- $\nu_{\text{f}} = 1$ sec$^{-1}$ and $J_{\text{s}} = 0.1 J$, one finds that the square-root law holds when the participation rate of the meta-order exceeds $3 \, 10^{-4}$, which is not unreasonable when compared with impact data. Interestingly the cross-over between a linear impact for small $Q$ and a square-root for larger $Q$ is consistent with the data presented by Zarinelli \emph{et al.} \cite{Zarinelli} (but note that the authors fit a logarithmic impact curve instead). \\

\subsection{Impact decay}

Regarding the decay impact for $t > T$, the problem to solve is that of Eqs.~\eqref{phi1}, \eqref{phi2} and \eqref{priceequal} only where  Eq.~\eqref{ratesequal} becomes:
\begin{eqnarray}
m_{\text{s}t}+m_{\text{f}t} &=& 0 \ .\label{ratessumzero}
\end{eqnarray}
The solution behaves asymptotically ($t\gg T$) to zero as $x_t \sim t^{-1/2}$ (see Appendix B). Given the results of Section~\ref{permimp} in the presence of finite memory agents,
the absence of permanent impact may seem counter-intuitive. In order to understand this feature of the double-frequency order book model in the limit $\nu_{\text{s}}\, T \rightarrow 0$, $\nu_{\text{f}}\, T\gg 1$, one can look at the stationary order book. As one moves away from the price the ratio of slow over fast volume fractions ($\phi_{\text{s}}/\phi_{\text{f}}$) grows linearly to infinity. Hence, the shape of the latent order book for $|x| \gg x^\star$ matches that of the infinite memory single-agent model originally presented by Donier \textit{et al.} \cite{DonierLLOB} (see Fig.~\ref{Obstat_multi}). This explains the mechanical return of the price to its initial value before execution, encoded in the slow latent order book. Note that in the limit of very small but finite $\nu_{\text{s}}$, the permanent impact is of order $\sqrt{\nu_{\text{s}}}$, as obtained in Section~\ref{permimp}.
\\

\subsection{The linear regime}

The regime of very small participation rates for which $m_0 \ll J_{\text{s}},J_{\text{f}}$ is also of conceptual interest. In such a case Eq.~\eqref{x1tspec} must be replaced with:
\begin{eqnarray}
x_{\text{s}t} &=&\frac{1}{\mathcal L_{\text{s}}} \int_0^t \textrm d \tau \, \frac{m_{\text{s}\tau}}{\sqrt{4\pi D (t-\tau)}}  \label{x1tspecbis} \ ,
%x_{\text{f}t} &=& \frac{\nu_{\text{f}}}{\lambda_{\text{f}}} \int_0^t \textrm d \tau \,  m_{\text{f}\tau}  \ .  \slabel{x2tspec} \label{x1tx2t}
\end{eqnarray}
which together with Eqs.~\eqref{x2tspec}, \eqref{ratesequal} and \eqref{priceequal} yields, in Laplace space (see Appendix B):
\begin{eqnarray}
\widehat m_{1p} &=&\frac1p \frac{m_0}{1+\sqrt{pt^\dagger}} \ , \label{m1p}
\end{eqnarray}
where $t^\dagger = (m_0/\pi J_{\text{s}}) t^\star$, with $t^\star$ defined in Eq. \eqref{m2}. For small times ($t \ll t^\dagger$) one obtains $m_{\text{s}t}= 2m_0 \sqrt{t/t^\dagger}$ while for larger times ($t^\dagger \ll t < T$), $m_{\text{s}t}=m_0[1-\sqrt{t^\dagger/(\pi t)}]$. Finally using again Eqs.~\eqref{x2tspec}, \eqref{ratesequal} and \eqref{priceequal} yields $x_t = (\nu_{\text{f}}/\lambda_{\text{f}})m_0 t$ for $t \ll t^\dagger$ and $x_t = (\nu_{\text{f}}/\lambda_{\text{f}})m_0\sqrt{t t^\dagger/\pi}$ for $t^\dagger \ll t < T$, identical in terms of scaling to the price dynamics observed in the case $J_{\text{s}} \ll m_0 \ll J_{\text{f}}$ discussed above. The asymptotic impact decay is identical to the one  obtained in that case as well.
\\
\section{Multi-frequency order book}
\label{densnusec}

The double-frequency framework {presented in Sec.~\ref{multifsec}} can be extended to the more realistic case of a continuous range of cancellation and deposition rates. Formally, one has to solve an infinite set of equations, labeled by the cancellation rate $\nu$: 
\begin{eqnarray}
\partial_t \phi_\nu = D \partial_{xx} \phi_\nu -\nu\phi_\nu +s_\nu(x,t)\ , \label{phinuc}
\end{eqnarray}
where $\phi_\nu(x,t)$ denotes the contribution of agents with typical frequency $\nu$ to the latent order book, and $s_\nu(x,t) = \lambda_\nu \,\sgn (x_{ \nu t}-x) + m_{\nu t}\delta(x-x_{ \nu t})$, with $\lambda_\nu =\mathcal L_{\nu}\sqrt{\nu D}$. Equation \eqref{phinuc} must then be completed with:
\begin{subeqnarray}
 \int_0^\infty \textrm d\nu \rho(\nu)  m_{\nu t}  &=&m_t \slabel{densnutaux} \\
x_{\nu t} &=&x_{t} \qquad \forall \nu \, , \label{densnutauxboth}
\end{subeqnarray}
where $\rho(\nu)$ denotes the distribution of cancellation rates $\nu$, and where we have allowed for an arbitrary order flow $m_t$. Solving exactly the above system of equations analytically is too ambitious a task. In the following, we present a simplified analysis that allows us to obtain an approximate scaling solution of the problem for a power law distribution of frequencies $\nu$.

\subsection{The propagator regime}
\label{diffusivitypuz}

We first assume, for simplicity, that the order flow $J_\nu$ is independent of frequency (see later for a more general case), and consider the case when $m_t \ll J$, $\forall t$. Although not trivially true, we assume (and check later on the solution) that this implies $m_{\nu t}\ll J$ $\forall \nu$, such that we can assume linear response for all $\nu$. Schematically, there are two regimes, depending on whether $t \gg \nu^{-1}$ -- in which case the corresponding density $\phi_\nu(x,t)$ has lost all its memory, or $t \ll \nu^{-1}$. In the former case the price trajectory follows Eq.~\eqref{x1tspecbis}, while in the latter case it is rather Eq.~\eqref{x2tspec} that rules the dynamics. One thus has:
\begin{subeqnarray}
\text{For } \nu t\ll 1 \quad  x_{ t} &=&\frac{1}{\mathcal L\sqrt{D}} \int_0^t \textrm d \tau \, \frac{ m_{\nu\tau}}{\sqrt{4\pi  (t-\tau)} }  \slabel{} \\
\text{For } \nu t\gg 1 \quad x_{ t} &=&\frac{\nu^{1/2}}{\mathcal L\sqrt{D}} \int_0^t \textrm d \tau \,  m_{\nu \tau}  \ .  \slabel{} \label{densnux}
\end{subeqnarray}
Inverting Eqs.~\eqref{densnux} and defining $\Psi(t) := 2/\sqrt{\pi t}$ yields (see Appendix B and in particular Eq.~\eqref{m1tx1point}):
\begin{subeqnarray}
\text{For } \nu t\ll 1 \quad  m_{\nu t}  &=& { \mathcal L\sqrt{D}} \int_0^t \textrm d \tau \, \Psi(t-\tau){\dot x_{ \tau }} \slabel{nupetit} \\
\text{For } \nu t\gg 1 \quad  m_{\nu t}  &=& \mathcal L \sqrt{D}{\nu^{-1/2}}   \dot  x_{ t} \ .  \slabel{nugrand} \label{nupetitgrand}
\end{subeqnarray}
Our approximation is to assume that $m_{\nu t}$ in Eq.~\eqref{densnutaux} is effectively given by Eq.~\eqref{nupetit} as soon as $\nu<1/t$ and by Eq.~\eqref{nugrand} when $\nu>1/t$ such that Eq.~\eqref{densnutaux} becomes:
\begin{eqnarray}
\int_0^{1/t} \textrm d\nu \rho(\nu)\bigg[\int_0^t \textrm d\tau  \Psi(t-\tau){\dot x_{ \tau }}\bigg]+ \int_{1/t}^\infty \textrm d\nu \rho(\nu)  \bigg[\nu^{-1/2} \dot  x_{ t} \bigg] &=&\frac {m_t}{\mathcal L \sqrt{D}} \ . \label{densnucentral}
\end{eqnarray}
Equation~\eqref{densnucentral} may be conveniently re-written as\footnote{We have implicitly defined the dimensionless functions $G(t) = \int_0^{1/t} \textrm d\nu \rho(\nu)$ and  $H(t) =t_{\textrm c}^{-1/2} \int_{1/t}^\infty \textrm d\nu \rho(\nu) \nu^{-1/2}$.} 
$\int_0^t \textrm d\tau \big[G(t) \Psi(t-\tau) + H(t)t_{\textrm c}^{1/2} \delta(t-\tau) \big] \dot x_\tau=  {m_t}/({\mathcal L \sqrt{D}})$.
Formally inverting the kernel $M(t,\tau):=\big[ G(t) \Psi(t-\tau) + H(t)t_{\textrm c}^{1/2} \delta(t-\tau) \big]$ then yields the price dynamics $\dot x_t$ as a linear convolution of the past order flow $m_{\tau \leq t}$. Note that when $m_t \to 0$, $\dot x_t$ is also small and hence, using Eqs.~\eqref{nupetitgrand}, all $m_{\nu t}$ are all small as well, justify our use of Eqs.~\eqref{densnux} for all frequencies.

%%%%%%%%%%%
%%%%%%%%%
\begin{figure}[t!]
\begin{center}
\resizebox{0.48\columnwidth}{!}{  \includegraphics{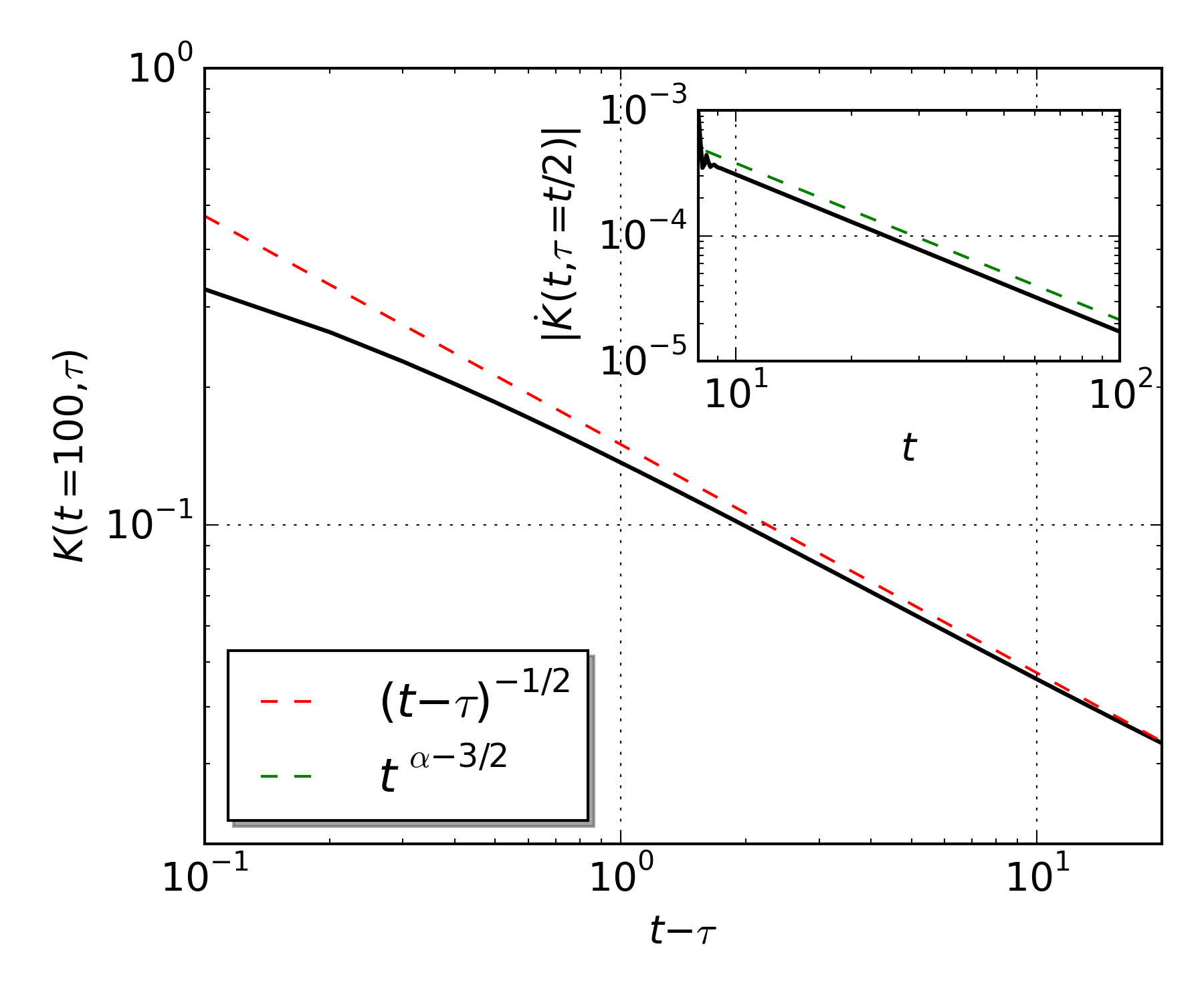}}
\end{center}
\caption{Numerical determination of the kernel $K(t,\tau):=M^{-1}(t,\tau)$, for $\alpha=0.25$. One clearly sees that $K$ decays as $(t-\tau)^{-1/2}$ at large lags. The inset 
shows that $K(t,t/2)$ behaves as $t^{\alpha - 1/2}$, as expected.}
\label{fig:kernel}      
\end{figure}
%%%%%%%%%
%%%%%%%%%%%

\subsection{Resolution of the ``diffusivity puzzle''}

Let us now compute the functions $G$ and $H$ for a specific power-law distribution $\rho(\nu)$ defined as:
\begin{eqnarray}
\rho(\nu)&=& Z \nu^{\alpha-1} e^{-\nu t_{\textrm c}} \ , \label{rhonudens}
\end{eqnarray}
where $\alpha>0$, $t_{\textrm c}$ is a high-frequency cutoff, and $Z=t_{\textrm c}^\alpha/\Gamma(\alpha)$.\footnote{Note that rigorously one should also introduce a low frequency cutoff $\nu_{\textrm{LF}}$ to ensure the existence of a stationary state of the order book in the absence of meta-order. Otherwise, $\langle\nu^{-1}\rangle=\infty$ when $\alpha \leq 1$ and the system does not reach a stationary state (see the end of Section \ref{llobrecall} and \cite{BenzaquenFLOB} for a further discussion of this point).}
For such a distribution, one obtains $G(t) = 1- \Gamma(\alpha,t_{\textrm c}/t)/\Gamma(\alpha)$ and $H(t) = \Gamma(\alpha-1/2,t_{\textrm c}/t)/\Gamma(\alpha)$. In the limit  $t\ll t_{\textrm c}, \ G(t)\approx 1$ and $H(t)\approx 0$. In the limit $t\gg t_{\textrm c}, \ G(t)\approx (t/t_{\textrm c})^{-\alpha}/[\alpha\Gamma(\alpha)]$,  and the dominant term in the first order expansion of $H(t)$ depends on whether $\alpha \lessgtr 1/2$. One has $H(t|_{\alpha<1/2})\approx 2 (t/t_{\textrm c})^{1/2-\alpha}/[\Gamma(\alpha)(1-2\alpha)]$ and $H(t|_{\alpha>1/2})\approx \Gamma(\alpha-1/2)/\Gamma(\alpha)$. Focussing on the interesting case $\alpha < 1/2$, one finds (see Fig.~\ref{fig:kernel}) that inversion of the kernel $M(t,\tau)$ is dominated, at large times, by the first term $G(t) \Psi(t-\tau)$. Hence, one finds in that regime:\footnote{Taking into account the $H(t)$ contribution turns out not to change the following scaling argument.}
\begin{eqnarray}
x_{t}&\approx&\frac{\alpha \Gamma(\alpha)}{\mathcal L t_{\textrm c}^{\alpha}  \sqrt{D} } \int_0^t \textrm d \tau \, 
\frac{m_\tau \tau^{\alpha}}{\sqrt{4\pi  (t-\tau)} } \ . \label{densnuprop}
\end{eqnarray}
Let us now show that this equation can lead to a diffusive price even in the presence of a long-range correlated order flow. Assuming that $\langle m_t m_{t'}\rangle \sim |t -t'|^{-\gamma}$ with $0 < \gamma < 1$ (defining a long memory process, as found empirically \cite{bouchaud2004fluctuations,bouchaud2008markets}), one finds from Eq. (\ref{densnuprop}) that the mean square price is given by:
\begin{eqnarray}
\langle x_t^2 \rangle \propto \iint_0^t \textrm d \tau \textrm d \tau' \frac{  \langle m_\tau m_{\tau'}\rangle {(\tau \tau')}^{\alpha}}{\sqrt{(t-\tau)(t-\tau')} } \ . 
\end{eqnarray}
Changing variables through $\tau \to tu$ and $\tau' \to tv$ easily yields $\langle x_t^2 \rangle \propto t^{1+2\alpha-\gamma}$. Note that the LLOB limit corresponds to a unique low-frequency for the latent liquidity. This limit can be formally recovered when $\alpha \to 0$. In this case, we recover the ``disease'' of the LLOB model, namely a mean-reverting, subdiffusive price  $\langle x_t^2 \rangle \propto t^{1-\gamma}$ for all values of $\gamma > 0$. Intuitively, the latent liquidity in the LLOB case is too persistent and prevents the price from diffusing. 
Imposing price diffusion, i.e. $\langle x_t^2 \rangle \propto t$ finally gives a consistency condition similar in spirit to the one obtained in \cite{bouchaud2004fluctuations}:
 \begin{eqnarray}
 \alpha&=& \frac{\gamma}{2} < \frac12 \ . \label{alphasgamma}
 \end{eqnarray}
Equation~\eqref{alphasgamma} states that for persistent order flow to be compatible with diffusive price dynamics, the long-memory of order flow must be somehow buffered by a long-memory of the liquidity, which makes sense. The present resolution of the diffusivity puzzle -- based on the memory of a multi-frequency self-renewing latent order book -- is similar to, but different from that developed in \cite{BenzaquenFLOB}. In the latter study we assumed the reassessment time of the latent orders to be fat-tailed, leading to a ``fractional'' diffusion equation for $\phi(x,t)$.

\subsection{Metaorder impact}

We now relax the constraint that $\lambda_\nu \propto \sqrt{\nu}$ and define $J_{\nu} := J_{\text{hf}} (\nu t_c)^{\zeta}$ 
with $\zeta>0$, meaning that HFT is the dominant contribution to trading, since in this case
\begin{eqnarray}
J&=& \int_{0}^\infty \textrm d\nu \rho(\nu) J_{\nu} =  J_{\text{hf}} \frac{\Gamma(\zeta+\alpha)}{\Gamma(\alpha)}. 
\end{eqnarray}
(The case $\zeta<0$ could be considered as well, but is probably less realistic).\\  

\begin{figure}[t!]
\begin{center}
\resizebox{0.48\columnwidth}{!}{  \includegraphics{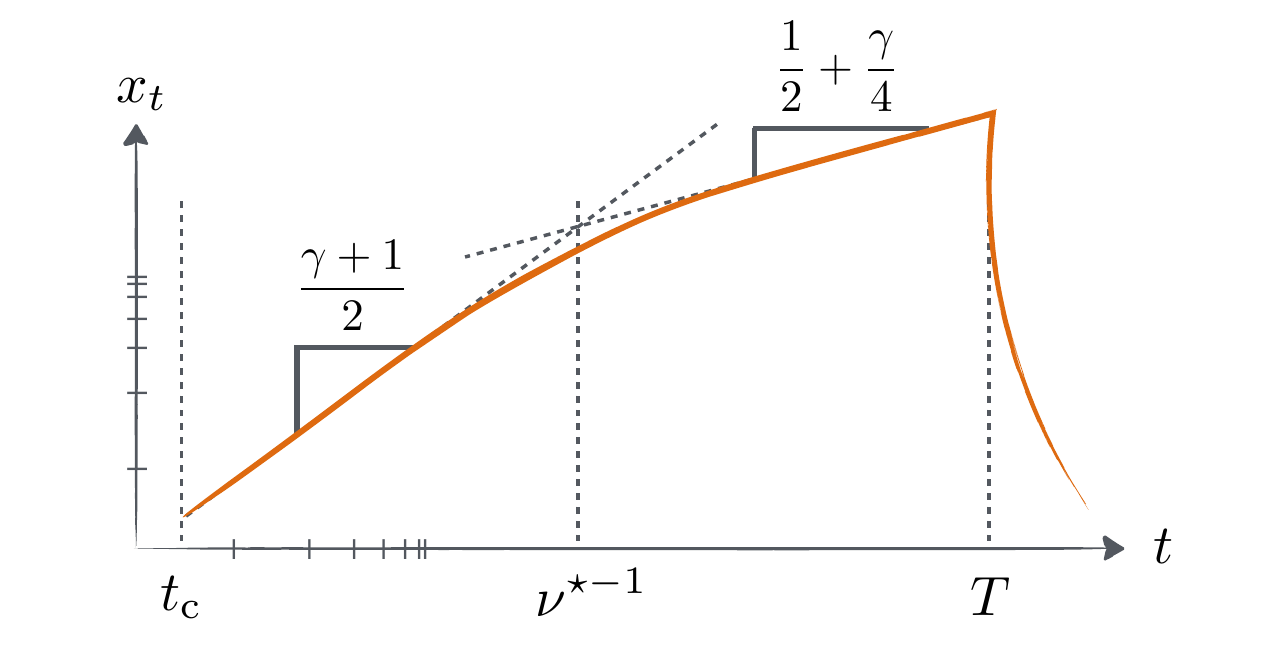}}
\end{center}
\caption{{Price trajectory during a constant rate metaorder execution within the multi-frequency order book model. For $\gamma=1/2$, the impact crosses over from a $t^{3/4}$ to a $t^{5/8}$ regime.}}
\label{pricetraj_multidens}      
\end{figure}

We consider a meta-order with constant execution rate $m_0 \ll J_{\text{hf}}$. Since $J_\nu$ decreases as the frequency decreases, there must exist a frequency $\nu^\star$ such that 
$m_0 = J_{\nu^\star}$, leading to $\nu^\star t_c = (m_0/J_{\text{hf}})^{1/\zeta}$. When $\nu \ll \nu^\star$, we end up in the non-linear, square-root regime where $m_0 \gg J_\nu$ and Eq. ~\eqref{x1tspec} holds. Proceeding as in the previous section, we obtain the following approximation for the price trajectory:
\begin{equation}
 G_\zeta(t) \bigg[\int_0^{t  }\textrm d\tau  \Psi(t-\tau)\dot x_{\tau}\mathds{1}_{\{ t\leq \nu^{\star -1}\}} + \frac{x_t \dot x_t}{2\sqrt D}  \mathds{1}_{\{ t>\nu^{\star -1}\}} \bigg]+t_c^{1/2} H_\zeta(t)  \dot  x_{t} =\frac{m_0 \sqrt{D}}{J_{\text{hf}}} \ . \label{densnucentral_s}
\end{equation}
where, in the limit $t\gg t_c$ and $\alpha + \zeta < 1/2$:
 \begin{subeqnarray}
G_\zeta(t)&:=& \int_0^{1/t} \textrm d\nu \rho(\nu)  (\nu t_c)^{\zeta} \approx  \left(\frac{t_c}{t}\right)^{\alpha+\zeta} \frac1{\Gamma(\alpha)(\alpha+s)}\\
H_\zeta(t)&:=& \int_{1/t}^\infty \textrm d\nu \rho(\nu) (\nu t_c)^{\zeta-1/2} \approx  \left(\frac{t_c}t\right)^{\alpha+\zeta-1/2} \frac1{\Gamma(\alpha)(1/2 - \alpha-s)}\ .
 \end{subeqnarray}
At short times $t \ll \nu^{\star -1}$, Eq.~\eqref{densnucentral_s} boils down to Eq.~\eqref{densnucentral} with $\alpha \rightarrow \alpha+\zeta$ and one correspondingly finds: 
\begin{equation}
x_t \propto  x_c \frac{m_0}{J_{\textrm{hf}}} \left(\frac{t}{t_c}\right)^{\frac12+\alpha+\zeta} \ ,
 \end{equation}
 where $x_c := \sqrt{Dt_c}$. For $t \gg \nu^{\star -1}$, the second term in Eq.~\eqref{densnucentral_s} dominates over both the first and the third terms, leading to a generalized 
square-root law of the form: 
\begin{equation}
x_t \propto x_c \sqrt{\frac{m_0}{J_{\text{hf}}}} \, \left(\frac{t}{t_c}\right)^{\frac{1+\alpha+\zeta}2} \ , 
\end{equation}
Compatibility with price diffusion imposes now that $\alpha + \zeta = \gamma/2$, which finally leads to (see Fig.~\ref{pricetraj_multidens}):
\begin{subeqnarray}
x_t &\propto&  x_c \frac{m_0}{J_{\textrm{hf}}} \,\left(\frac{t}{t_c}\right) ^{\frac{1+\gamma}{2}}, \quad {\text{when}} \quad  t \ll t_c \left(\frac{J_{\text{hf}}}{m_0}\right)^{1/\zeta} \\
x_t &\propto& x_c \sqrt{\frac{m_0}{J_{\text{hf}}}}\, \left(\frac{t}{t_c}\right)^{\frac{2+\gamma}{4}}, \quad {\text{when}} \quad t \gg t_c \left(\frac{J_{\text{hf}}}{m_0}\right)^{1/\zeta} \ .
\end{subeqnarray}
In the latter case, setting $\gamma = 1/2$ and $Q = m_0 T$, one finds an impact $I_Q:=x_T$ behaving as\footnote{{Note that $5/8\approx 0.6$ is very close close to the empirical impact results reported by Almgren \emph{et al.} and Brockmann \emph{et al.} \cite{Almgren2005,Brockmann2015} in the case of equities, for which $\gamma$ is usually close to 1/2.} } $Q^{5/8}$ as soon as $Q > \upsilon (J_{\text{hf}}/m_0)^{(1-\zeta)/\zeta}$, where we have introduced an elementary volume $\upsilon := J_{\text{hf}} t_c$, which is the volume traded by HFT during their typical cancellation time.

%%%%%%%%%%%
%%%%%%%%%
%%%%%%%%%
%%%%%%%%%%%

\section{Conclusion}
\label{concl}

In this work, we have extended the LLOB latent liquidity model \cite{DonierLLOB} to account for the presence of agents with different memory timescales. This has allowed us to overcome several conceptual and empirical difficulties faced by the LLOB model. We have first shown that whenever the longest memory time is finite (rather than divergent in the LLOB model), a permanent component of impact appears, even in the absence of any ``informed'' trades. This permanent impact is {\it linear} in the traded quantity {  and independent of the trading rate}, as imposed by no-arbitrage arguments. We have then shown that the square-root impact law holds provided the meta-order participation rate is large compared to the trading rate of ``slow'' actors, which can be small compared to the total trading rate of the market -- itself dominated by high-frequency traders. In the original LLOB model where all actors are slow, a square-root impact law independent of the participation rate only holds when the participation rate is large compared to the total market rate, which is not consistent with empirical data. Finally, the multi-scale latent liquidity model offers a new resolution of the diffusivity paradox, i.e. how an order flow with long-range memory can give rise to a purely diffusive price. We show that when the liquidity memory times are themselves fat-tailed, mean-reversion effects induced by a persistent order book can exactly offset trending effects induced by a persistent order flow. \\

We therefore believe that the multi-timescale latent order book view of markets, encapsulated by Eqs.~\eqref{phinuc} and \eqref{densnutauxboth}, is rich enough to capture a large part of the subtleties of the dynamics of markets. It suggests an alternative framework to build agent based models of markets that generate realistic price series, that complement and maybe simplify previous attempts \cite{Toth2011,mastromatteo2014agent}. A remaining outstanding problem, however, is to reconcile the extended LLOB model proposed in this paper with some other well known ``stylized facts'' of financial price series, namely power-law distributed price jumps and clustered volatility. We hope to report progress in that direction soon. Another, more mathematical endeavour is to give a rigorous meaning to the multi-timescale reaction model underlying Eqs.~\eqref{phinuc} and \eqref{densnutauxboth} and to the approximate solutions provided in this paper. It would be satisfying to extend the no-arbitrage result of Donier et al. \cite{DonierLLOB}, valid for the LLOB model, to the present multi-timescale setting.  \\

We thank J. Bonart, A. Darmon, J. de Lataillade, J. Donier, Z. Eisler, A. Fosset, S. Gualdi, I. Mastromatteo, M. Rosenbaum and B. T\'oth for extremely fruitful discussions.

\clearpage

\section*{Appendix A}
\label{sec:Appendix1}

We here provide the calculations that link Eq.~\eqref{alphaz0z1} and Table~\ref{tableimpact} during a meta-order execution ($t\leq T$); the impact decay computations ($t>T$) are given and discussed in Section \ref{permimp}.\\
 
In the limit of slow execution of the meta-order, one has ${(x_t-x_\tau)^2}\ll {{4D(t-\tau)}}$ such that Eq.~\eqref{mastereq} together with Eq.~\eqref{priceeq} becomes:
\begin{eqnarray}
0 &=&  \phi^\mathrm{st}(x_t)e^{-\nu t} + \int_0^{t} \text d \tau\, \frac{m_0}{\sqrt{4\pi D(t-\tau)} }  e^{-\nu(t-\tau) }    -{2 \lambda}\int_0^{t } \text d \tau \,   \frac{x_t-x_\tau}{\sqrt{4\pi D(t-\tau)} }  e^{-\nu(t-\tau)} \ .\label{slowshort}
\end{eqnarray}
Interestingly, slow and short execution is only compatible with small meta-order volume\footnote{Equivalently, rapid and long execution is only consistent with large meta-order volume (combining $m_0\gg J$ and $\nu T\gg 1$ implies $m_0 T \gg J \nu^{-1}$).} (indeed, combining $m_0\ll J$ and $\nu T\ll 1$ implies $m_0 T \ll J \nu^{-1}$). Thus for  slow and short execution, using the linear approximation  $\phi^\mathrm{st}(x_t)=-\mathcal L x_t$ and letting Eq.~\eqref{alphaz0z1} into Eq.~\eqref{slowshort} yields:
\begin{subeqnarray}
0&=&  -\mathcal L \alpha z^0_t +m_0\sqrt{\frac t{\pi D}}\slabel{slowshort0th} \\
0&=&   -\mathcal L\sqrt{\nu} z^1_t - 2\lambda \int_0^t \mathrm d \tau\, \frac{z_t^0-z_\tau^0}{\sqrt{4\pi D(t-\tau)}}\ .\slabel{slowshort1st}
\end{subeqnarray}
Equation \eqref{slowshort0th} yields $\alpha = m_0/(\mathcal L\sqrt{\pi D})$ and $z_t^0=\sqrt{t}$, and it follows from  Eq.~\eqref{slowshort1st} that $z_t^1 = - kt$ where $k=\sqrt{4/\pi} - \sqrt{\pi/4}$. \\

In the limit of fast execution, one has ${(x_t-x_\tau)^2}\gg {{4D(t-\tau)}}$ such that the meta-order term can be approximated through the saddle point method. Letting $x_\tau \approx x_t- (t-\tau)\dot x_t$ into the price equation now yields:
\begin{eqnarray}
0 &=&  \phi^\mathrm{st}(x_t)e^{-\nu t} + \int_0^{t} \text d \tau\, m_0 \frac{e^{-\frac{\dot x_t^2(t-\tau)}{4D}}}{\sqrt{4\pi D(t-\tau)} }  e^{-\nu(t-\tau) } 
 -{ \lambda}\int_0^{t } \text d \tau \,  e^{-\nu(t-\tau)} \ .\label{fastshort}
\end{eqnarray}
Letting $u=t-\tau$ and given ${4D}/{\dot x_t^2}\ll t$ such that $\int_0^t \mathrm du  \approx \int_0^\infty \mathrm du$, Eq.~\eqref{fastshort} becomes:
\begin{eqnarray}
0 &=&  \phi^\mathrm{st}(x_t)e^{-\nu t} + \frac{m_0}{\sqrt{\dot x_t^2+4D\nu}}  +\frac{ \lambda}\nu\left(  e^{-\nu t}-1\right) \, .\label{fastshortbis}
%\int_0^{t} \text d u\,  \frac{ m_0e^{-\frac{\dot x_t^2u}{4D}}}{\sqrt{4\pi Du} }  e^{-\nu u } 
\end{eqnarray}
For short execution with small meta-order volume (we use $\phi^\mathrm{st}(x_t)=-\mathcal L x_t$), letting  Eq.~\eqref{alphaz0z1} into Eq.~\eqref{fastshortbis} yields:
\begin{subeqnarray}
0&=&  -\mathcal L \alpha z^0_t + \frac{m_0}{\alpha |\dot z_t^0|} \slabel{fastshortbis0th} \\
0&=& -\mathcal L \alpha \sqrt{\nu}z^1_t  - \frac{\sqrt{\nu}m_0}{\alpha}\frac{\dot z_t^1}{ (\dot z_t^0)^2} -\lambda t  \ .\slabel{fastshortbis1st}
\end{subeqnarray}
Equation \eqref{fastshortbis0th} yields $\alpha = \sqrt{{2m_0}/{\mathcal L}}$ and $z_t^0=\sqrt{t}$, and thus   Eq.~\eqref{fastshortbis1st} becomes $\dot z_t^1 + {z_t^1}/({2t}) = - \frac12\sqrt{J/({2m_0})}$. It follows that $z_t^1 =  - \frac t3\sqrt{J/(2m_0)} $.
For a fast, short and large meta-order, $x_t$ is expected to go well beyond the linear region of the order book such that in a hand-waving static approach (consistent with fast and short execution) one can match $m_0 t$ and the area of a rectangle of sides $x_t$ and $\lambda\nu^{-1}$ (see Fig.~\ref{Obstat}). Letting $x_t=b t$ yields $b = m_0\nu /\lambda$. Note that this result can be recovered by letting $x_t=b t$ and $\phi^\mathrm{st}(x_t)=-\lambda \nu^{-1}$ into Eq.~\eqref{fastshortbis}. Indeed, at leading order one obtains: 
\begin{eqnarray}
0&=&  -\frac{\lambda}\nu + \frac{m_0}{ |\dot x_t|}  \ ,\label{}
\end{eqnarray}
from which the result trivially follows.\\

For long execution ($\nu T\gg1$) the memory of the initial book is rapidly lost and one expects Markovian behaviour.  Letting again $x_t=b t$ into the price equation and changing variables through $\tau =t(1-u)$ yields:
\begin{eqnarray}
0 &=&  m_0 \sqrt{t} \int_0^1 \text d u\,\frac{  e^{-\frac{b ^2tu}{4D}}}{\sqrt{4\pi Du}}  e^{-\nu t u} 
   -\lambda\int_0^1 \text d u \textstyle\,e^{-\nu t u}  \, \erf  \sqrt{\frac{b^2tu}{4D}}   \nonumber  \\ &=&  \left(m_0 - \frac{\lambda b }{\nu}\right)\frac{1}{\sqrt{b ^2+4D\nu }}\, \erf \,\textstyle  \sqrt{\left( \frac{b ^2}{4D}+\nu\right) t  }\ . \label{longall}
\end{eqnarray}
Interestingly, Eq.~\eqref{longall} yields $b = m_0\nu /\lambda$ (regardless of execution rate and meta-order size), which is exactly the result obtained above in the case of fast and short execution of a large meta-order but for different reasons.

\section*{Appendix B }
\label{sec:Appendix2}

We here provide the calculations underlying  the double-frequency order book model presented in Section~\ref{multifsec}. In particular {for the case $J_{\text{s}}\ll m_0\ll  J_{\text{f}}$}, Eqs.~\eqref{x1tx2t} are obtained as follows. In the limit of large trading intensities the saddle point methods (as detailed in Appendix A) can also be applied to the case of nonconstant execution rates (one lets $m_\tau \approx m_t$ about which the integrand is evaluated, see \cite{DonierLLOB}), in particular one obtains (equivalent to Eq.~\eqref{fastshortbis1st}):
\begin{eqnarray}
\mathcal L_{\text{s}} x_{\text{s}t}|\dot x_{\text{s}t}|&=&   {m_{\text{s}t}}\ , \label{} 
\end{eqnarray}
which yields Eq.~\eqref{x1tspec}.  For the rapid agents ($\nu_{\text{f}}T\gg 1$) we must consider the case of long execution. In particular, an equation tantamount to Eq.~\eqref{longall} can also be derived in the case of nonconstant execution rates. Proceeding in the same manner, one easilly obtains:
\begin{eqnarray}
0 &=&  \left( m_{\text{f}t} - \frac{\lambda_{\text{f}} \dot x_{\text{f}t} }{\nu_{\text{f}}} \right) \frac{1}{\sqrt{\dot x_{\text{f}t} ^2+4D\nu_{\text{f}} }}\, \erf \,\textstyle  \sqrt{\left( \frac{\dot x_{\text{f}t}^2}{4D}+\nu_{\text{f}}\right) t  }\ , \quad\quad \label{}
\end{eqnarray}
which yields ${ \dot x_{\text{f}t} } = m_{\text{f}t}\nu_{\text{f}} /\lambda_{\text{f}}$ and thus Eq.~\eqref{x2tspec}. Then, as mentioned in Section~\ref{multifsec}, the asymptotic impact decay is obtained from Eqs.~\eqref{phi1}, \eqref{phi2} and \eqref{priceequal} 
only where for $t>T$ we replace Eq.~\eqref{ratesequal} with Eq.~\eqref{ratessumzero}. Using Eq.~\eqref{mastereq} together with Eq.~\eqref{priceeq}  in the limit $\nu_{\text{s}}T\rightarrow 0$, and $\nu_{\text{f}}T\gg 1$ together with \eqref{priceequal} yields ($t>T$):
\begin{subeqnarray}
\mathcal L_{\text{s}} x_t &=& \int_0^T \!\!\!\! +\! \int_T^t  \textrm d\tau \frac{ m_{\text{s}\tau} }{\sqrt{4\pi D(t-\tau)}}    \\
0&=& \int_0^T  \!\!\!\! +\! \int_T^t  \textrm d\tau \frac{e^{-\nu_{\text{f}}(t-\tau)}}{\sqrt{4\pi D (t-\tau)}}\big[   m_{\text{f}\tau} - 2\lambda_{\text{f}}(x_t-x_\tau) \big] \, . \quad \quad \label{decaymultifreq}
\end{subeqnarray}
Asymptotically ($t\gg T$) the system of Eqs.~\eqref{decaymultifreq} becomes: 
\begin{subeqnarray}
\mathcal L_{\text{s}} x_t &=& \int_0^T   \frac{ m_{\text{s}\tau} \textrm d\tau}{\sqrt{4\pi D(t-\tau)}} +\int_T^t    \frac{m_{\text{s}\tau}  \textrm d\tau}{\sqrt{4\pi D(t-\tau)}}   \slabel{decm1}  \\
0&=&\int_0^t \textrm d\tau \frac{e^{-\nu_{\text{f}}(t-\tau)}}{\sqrt{4\pi D (t-\tau)}}\left[   m_{\text{f}\tau} - 2\lambda_{\text{f}}(x_t-x_\tau) \right]  . \quad \quad \slabel{decm2} \label{decaymultifreqsimp}
\end{subeqnarray}
We expect the asymptotic impact decay to be of the form $x_t = x_\infty + B/\sqrt{t}$. In addition Eq.~\eqref{decm2} indicates that $m_{\text{f}t} \sim \dot x_t$. We thus let $m_{\text{s}t}=-m_{\text{f}t}=C/t^{3/2}$. Injecting into Eq.~\eqref{decm1} yields $x_\infty = 0$ (no permanent impact) and:
\begin{eqnarray}
\frac{\mathcal L_{\text{s}} B}{\sqrt{t}} &=&\frac1{\sqrt{t}} \left[ \frac{m_0f_{T}}{\sqrt{4\pi D}}  +\frac{C}{\sqrt{\pi DT}} \right] \ , \label{l1Bsqrtt}
\end{eqnarray}
where $ f_T=T$ if $t^\star\ll T$ and $f_T = T^2/(3t^\star)$ if $ t^\star \gg T$.
On the other hand, letting $u=t-\tau$ in Eq.~\eqref{decm2} and using $x_t-x_s \approx (t-s)\dot x_t$ yields at leading order:
\begin{eqnarray}
0&=&\int_0^\infty \textrm du \,\frac{e^{-\nu_{\text{f}}u}}{\sqrt{u}}\left[ -\frac C{t^{3/2}} + \frac{\lambda_{\text{f}}B u}{t^{3/2}} \right]  
=  \sqrt{\frac{\pi}{\nu_{\text{f}}t^{3}}}\left[  -C + \frac{\lambda_{\text{f}} B}{\nu_{\text{f}}}\right]\label{} \ ,
\end{eqnarray}
which combined with Eq.~\eqref{l1Bsqrtt} easily leads to the values of $B$ and $C$.\\ %

 For the case $m_0\ll J_{\text{s}},J_{\text{f}}$, the calculations are slightly more subtle. Inverting Eq.~\eqref{x1tspecbis} in Laplace space yields:
\begin{eqnarray}
 m_{\text{s}t} &=&2{\mathcal L_{\text{s}}} \sqrt{D} \int_0^t \textrm d \tau \, \frac{ \dot x_{\text{s}\tau}}{\sqrt{\pi (t-\tau)} }  \label{m1tx1point} \ .
\end{eqnarray}
One can easily check this result by re-injecting Eq.~\eqref{m1tx1point} into Eq.~\eqref{x1tspecbis}. In turn, inverting Eq.~\eqref{x2tspec} is straightforward and yields $
 m_{\text{f}t} =({\lambda_{\text{f}}}/{\nu_{\text{f}}})\dot x_{\text{f}t}$. 
Injecting $\dot x_{\text{s}t}=\dot x_{\text{f}t}$ into Eq.~\eqref{m1tx1point} and using Eq.~\eqref{ratesequal} yields:
\begin{eqnarray}
 m_{\text{s}t} &=&\frac 1{\sqrt{t^\dagger}} \int_0^t \textrm d \tau \, \frac{ m_0- m_{\text{s}\tau}}{\sqrt{\pi(t-\tau)} }  \label{} \ ,
\end{eqnarray}
{which can be written as:
\begin{eqnarray}
\int_0^t \textrm d \tau \, { m_{\text{s}\tau}}\Phi(t-\tau) =  2m_0 \sqrt{t}\ ,   \quad \text{with} \ \Phi(t) := {\delta(t)}{\sqrt{\pi t^\dagger}} +\frac{\theta(t)}{\sqrt{t}} \ .  \label{realtolapm1} 
\end{eqnarray}
Taking the Laplace transform of Eq.~\eqref{realtolapm1} one obtains $\widehat \Phi(p) \widehat m_{sp}=m_0\sqrt{\pi}/p^{3/2}$ with $\widehat \Phi(p)=\sqrt{\pi t^\dagger}+\sqrt{\pi/p}$,
which in turn yields Eq.~\eqref{m1p}.}

\clearpage

\bibliographystyle{iopart-num}
\bibliography{bibs}

\end{document}